\documentclass{article}

\usepackage{graphicx}  
\usepackage{amsmath}   
\usepackage[compress]{cite}
\usepackage{amssymb}   
\usepackage{bm} 

\usepackage{mathrsfs}

\usepackage[active]{srcltx}

\def\eq#1{{Eq.~(\ref{#1})}}

 \def\g{\sqrt{-g}\,}

\def\gm#1{{\mathcal{G}^a[#1]}}
\def\mm#1{{\mathcal{P}^a[#1]}}

\newcommand{\LL}{Lanczos-Lovelock }

\title{Momentum density of spacetime and the gravitational dynamics}

\author{T. Padmanabhan\\
IUCAA, Post Bag 4, Ganeshkhind,
 Pune - 411 007, India.\\
email: paddy@iucaa.ernet.in}

\date{ }

\begin{document}

\maketitle

\begin{abstract}
I introduce a covariant four-vector $\gm v$, which can be interpreted as  the momentum density attributed to the spacetime geometry by an observer with velocity $v^a$,  and describe its properties: 
(a) Demanding that the total momentum of matter plus geometry is conserved for all observers, leads to the gravitational field equations.
Thus, how matter curves spacetime is entirely determined by this  principle of momentum conservation. 
(b)  The  $\gm v$ can be related to the gravitational Lagrangian in a manner similar to 
 the usual definition of Hamiltonian in, say, classical mechanics.
(c) Geodesic observers in a spacetime will find that the conserved total momentum vanishes on-shell.
(d) The on-shell, conserved, \textit{total} energy in a region of space, as measured by comoving observers,  will be equal to the total heat energy of the boundary surface.  
(e) The off-shell \textit{gravitational energy} in a region will be the sum of the ADM energy in  the bulk plus the thermal energy of the boundary. 
These results suggest that $\gm v$ can be a useful physical quantity to probe the gravitational theories.
\end{abstract}

Matter possesses an energy-momentum tensor $T^a_b$. Given the validity of $\partial_a T^a_b =0$ in the freely falling frames, where special relativity holds, and the  principle of general covariance and principle of equivalence, one can impose the condition $\nabla_a T^a_b =0$ in any arbitrary curved spacetime. In a generic situation, $\nabla_a T^a_b =0$ will lead to the equations of motion for matter in the given spacetime. So, ``how geometry makes the matter move'' is encoded in the generalized conservation law $\nabla_a T^a_b =0$. 

Can we have a similar principle to determine ``how matter curves the  geometry''? That is, can we get the gravitational field \footnote{The signature is $(-,+,+,+)$. We use units with $\hbar = c= (16\pi G) =1$ so that Einstein's equations become $G^a_b = (1/2)T^a_b$. Latin letters run through  $0-3$ and Greek letters run through $1-3$.} equation
 $2G^a_b = T^a_b$ from a \textit{generally covariant} conservation law? As is well known, one cannot define a generally covariant local energy-momentum tensor $t^a_b$ for gravity. It is possible to obtain field equations from a relation like  $\partial_a (t^a_b +T^a_b)=0$ with many different \cite{mtw} pseudo-tensors $t^a_b$. But we are looking for a generally covariant law. 

I will now show how this can be done.
The key idea is to shift attention from the energy-momentum tensor to  the momentum vector. Even for normal matter, the momentum density $P^a \equiv -T^a_b v^b$ can be defined \textit{only} by using an additional vector field $v^a$, say, the velocity  of an observer. That is, while $T^a_b$ can be expressed entirely in terms of matter fields and geometry, the corresponding four-momentum $P^a$ associated with matter is different for different observers; and 
requires an additional vector field for its definition.  It will, therefore, be only natural if different observers attribute different momentum density $\gm v$ for the same spacetime geometry as well. Further, the momentum of the matter field $-T^a_b v^b$ defined with the velocity $v^b$ of an arbitrary observer is \textit{not} conserved because $\nabla_a(T^a_b v^b) =T^a_b  \nabla_av^b$ is, in general, non-zero. 

These facts suggest that we could restore the momentum conservation by adding the matter momentum to the momentum attributed to the spacetime geometry by the same observer. That is, we expect the correct physical law of nature to be expressed in the form 
\begin{equation}
\nabla_a \left( \gm v + P^a[v]\right) \equiv \nabla_a \mm v =0
\label{momcon}
\end{equation} 
where $\gm v $ is the momentum density attributed to a spacetime geometry by an observer with velocity $v^a$.
Just as  $\nabla_a T^a_b =0$ tells us  ``how geometry makes the matter move'', we expect $\nabla_a \mm v =0$ to tell us ``how matter curves the  geometry''.  In short, \textit{gravitational field equations represent the conservation law for the  total momentum.}

 I will now give a definition of $\gm q$ and show how the gravitational field equations arise from \eq{momcon}. (It turns out to be useful to define this quantity for an arbitrary vector field $q^a$ rather than for those with $v^2=-1$, which can arise as a special case.) Having done that, I will explain why this definition is natural and how it provides deeper insights into the nature of gravity.

I will begin with the description of the gravitational field in terms of the variables (see \cite{KBP,grtp} for a detailed discussion of these variables): 
\begin{equation}
f^{ab}\equiv \g\, g^{ab}; \qquad N^a_{bc} \equiv - \Gamma^a_{bc} + \frac{1}{2} \left(\Gamma^d_{bd} \delta^a_c + \Gamma^d_{cd} \delta^a_b \right) 
\label{b1}
\end{equation} 
instead of the usual pair $(g_{ab},\Gamma^i_{jk})$. The spacetime momentum $\gm q$ associated with a  vector field $q^a$ is defined in terms of the variables in \eq{b1} by\footnote{This definition will lead to Einstein's theory. There is a natural extension of all the results in this paper to \LL\ models which will be presented elsewhere.}
\begin{equation}
\g\, \gm q \equiv -\left[\left( \g\, R\right) q^a + f^{ij} \pounds_q N^a_{ij}\right]
\label{a3}
\end{equation}
where the Lie derivative of $N^i_{jk}$,  defined in terms of the Lie derivative of the connection
$
\pounds_q\Gamma^a_{bc}=\nabla_b \nabla_c q^a+R^a_{\phantom{a}cmb}q^m
$, is generally covariant.
To prove that \eq{momcon} --- along  with $\gm q$ defined by \eq{a3} --- implies the field equations, we will proceed as follows. 
From the anti-symmetric part of the derivative  $\nabla^{[l} q^{m]} \equiv J^{lm}$ of any vector field $q^a$ we immediately get a conserved current $J^i\equiv \nabla_k J^{ik}$. Manipulating the derivatives, it is easy to express \cite{grtp, mpla} this current\footnote{This \textit{is} indeed the \textit{off-shell}, identically conserved, Noether current associated with  $q^a$. I stress that it can be obtained purely from a differential geometric identity \textit{without mentioning the action principle for gravity or any diffeomorphism invariance!}\cite{grtp}.
In normal units, the left hand side should be multiplied by $16\pi G$ which we have set to unity.} as:  
\begin{equation}
 J^a[q] = \nabla_b J^{ab} [q] = 2 R^a_b q^b + g^{ij} \pounds_q N^a_{ij}
\label{noe}
\end{equation}
Therefore 
$\gm v = 2 G^a_b v^b-J^a[v]$. 
On using this relation, the definition $P^a[v] \equiv -T^a_b v^b$ and the identities, $\nabla_a T^a_b =0,\  \nabla_a G^a_b =0$, \eq{momcon} reduces to the relation 
\begin{equation}
\left( 2 G^a_b - T^a_b\right) \, \left( \nabla_av^b \right) = 0
\label{a6}
\end{equation} 
We now demand that \eq{momcon} should hold for all observers and hence \eq{a6} should hold for all $v^a$ at all events. Since $\nabla_a v^b$ at any given event is arbitrary, this leads\footnote{The fact that $v^a v_a = -1$ does not affect the argument. One can see this more formally by writing \eq{a6} in a local inertial frame near the origin (with $\nabla_a v^b =\partial_a v^b$) and taking $v^a=q^a/(-q_iq^i)^{1/2}$ where $q^a$ is an arbitrary timelike vector.  Using the Taylor series expansion $q^b(x) = q^b(0) + M^b_{\phantom{b}c}(0) x^c+\mathcal{O}(x^2)$, it is easy to see that we have sufficient freedom in the choice of $ q^b(0),  M^b_{\phantom{b}c}(0)$ to validate the above argument.} to the field equations $2G^a_b = T^a_b$. 
Thus, with the definition of gravitational momentum in \eq{a3}, we can express gravitational field equations as a conservation law. 

I will now explore the consequences of this conservation law and argue that the definition in \eq{a3} is quite natural and useful.

To begin with, it is conceptually  rather elegant to describe  ``how geometry tells the matter to move'' \textit{and} ``how matter tells the  geometry to curve'' in terms of the two principles $\nabla_a T^a_b =0$ and $\nabla_a \mm v =0$. While the conservation (or otherwise!) of the matter energy momentum tensor is widely discussed in literature, the fact that matter \textit{momentum} is not conserved (viz. $\nabla_a P^a[v] \neq 0$, in a general spacetime) does not seem to have received much attention.  
\textit{Our principle shows that the momentum conservation law is indeed restored once we take the spacetime momentum into account.} Or, rather, matter momentum alone was not conserved because we did not add to it the momentum of the spacetime geometry produced by the matter.

Second, unlike $\nabla_a T^a_b =0$,  \eq{momcon} describes a genuine conservation law with an associated conserved charge that can be interpreted as the total energy.
From 
$\gm q = 2 G^a_b q^b-J^a[q]$ we find that the total, conserved, on-shell momentum $\mm q=-J^a[q]$ is essentially the Noether current. (The negative sign is due to our signature.)
Since 
the conservation of $J^a[q]$ can \textit{also} be related to the diffeomorphism invariance of the action under $x^a\to x^a+q^a$, it seems natural to see it emerge as the total momentum.

Third, the structure of $\gm q$ is similar to the usual definition of Hamiltonian from the Lagrangian, generalized suitably for Einstein's theory. In classical mechanics, given a Lagrangian $L_q(\dot q,q)$ which leads to second-order equations of motion when $\delta q=0$ at the boundary, one can construct another (`momentum-space') Lagrangian $L_p(\ddot q,\dot q,q) \equiv L_q - d(pq)/dt$ which will lead to the same \textit{second-order} equations of motion --- in spite of the fact that  $L_p(\ddot q,\dot q,q)$ depends on $\ddot q$ --- when $\delta p=0$ at the boundary \cite{ayan}. The standard Hamiltonian is expressible in terms of $L_q$ or $L_p$ as:
\begin{equation}
H = - L_q + p \dot q  = - (L_p + q \dot p)
\label{a7} 
\end{equation} 
In terms of the variables $(q,p)\Leftrightarrow (f^{ab},N^i_{jk})$, gravity can be similarly \cite{KBP} described by either of the two Lagrangians, $L_f$ (which leads to the field equations when $\delta f^{ab}=0$ at the boundary, usually called the $\Gamma^2$ Lagrangian) and $L_N=\sqrt{-g}R$ (which leads to the same field equations when $\delta N^a_{bc}=0$  at the boundary). The $L_N$ and $L_f$ are related by:
\begin{equation}
L_N = L_f - \partial_c (f^{ab}N^c_{ab})  
= \g R                                              
\end{equation} 
where $N^a_{bc}=\partial(\sqrt{-g}R)/\partial(\partial_af^{bc})$ is the  momentum conjugate to $f^{ab}$.
 In fact, the variation of $L_N=\g R$ can be written as:
\begin{equation}
 \delta(\sqrt{-g}R)=  R_{ab}\delta f^{ab}-\partial_{c}[f^{ik}\delta N^{c}_{ik}]
 =\g[G_{ab}\delta g^{ab}-\nabla_{c}(g^{ik}\delta N^{c}_{ik})]
 \label{varRNf}
\end{equation}
showing that $\delta N^{c}_{ik}=0$ at the boundary will lead to the equations of motion. That is, the  Lagrangian $\sqrt{-g}R$ has  a momentum space structure.
 Our definition of the gravitational four momentum in \eq{a3} is a direct generalization of (the second equality in) \eq{a7} with  the pair  $(f^{ab},N^i_{jk})$  which arises in \eq{varRNf} replacing  $(q,p)$, and the Lie derivative of $\pounds_v N^i_{jk}$ ---which is generally covariant ---  replacing $\dot p$, leading to the definition in \eq{a3}.
(The multiplication of  $L$ by $v^a$  gets us the four vector index.) 
This suggests that the definition has a natural relationship with the Hilbert action. 

Since our starting point is the conservation law in \eq{momcon}, we could have added to $\gm q$ any conserved current and still obtained the field equation $2G^a_b=T^a_b$. (In fact, one cheap trick will be to define the gravitational momentum as simply $2G^a_bq^b$, which would have directly lead
to \eq{a6} but the total on-shell momentum will always be zero!). While adding an arbitrary conserved current does not affect the derivation of the field equations, it will certainly change (a) the value of the conserved charge and (b) the relation of the gravitational momentum to the Hilbert action. The above discussion motivates why our definition is natural. In addition to having the a structure similar to the one in \eq{a7}, it also identifies the total momentum as the Noether current $J^a[q]$. 

The value of the corresponding conserved charge will, of course, depend on the vector field $q^a$ chosen for the definition. But since the properties of the Noether current are well-known from the previous works \cite{grtp, mpla}, we will be able to obtain the corresponding results easily. I will describe the results for two natural choices.

The first class of observers we can look at are the geodesic observers in the synchronous frame, which can be introduced in any local region of any spacetime. These observers have $u_a=-\nabla t(x)$ and hence $J^{ab}[u]$ and $J^a[u]$ vanish for these observers \cite{mpla}. So these geodesic observers will find the total four-momentum $\mm u$ to vanish. We know that geodesic observers notice the absence of local gravitational field but, of course, experience the tidal effect of gravity. The fact that $\mm u =0$ for these observers provides another nice characterization of the synchronous frame.\footnote{The Noether current $J^a[q]$ is invariant under the `gauge transformation' $q^a\to q^a+\partial^af$. In fact all the relevant algebra is identical to electrodynamics in curved spacetime with $q^i\Leftrightarrow A^i, J^{mn}\Leftrightarrow F^{mn},
J^a\Leftrightarrow j^a$ where $j^a$ is an electromagnetic current sourcing $A^i$. Any vector $q^i$ which satisfies source-free Maxwell's equations in a given metric will have zero Noether current. Further, using the gauge freedom we can always set $\nabla_iq^i=0$ and for such vector fields \eq{a6} will give the trace-free Einstein's equations, which will lead to Einstein's equations with the cosmological constant arising as an integration constant. I hope to revisit this idea in a later work. } 

The second set of observers are those moving normal to the spacelike hypersurfaces in a given foliation. To explore this situation, let us introduce
an arbitrary ($1+3$) foliation based on a time function $t(x^a)$, with the unit normal $u_a(x^i)\propto \nabla_at$. This leads to the  (1+3) split of the metric $g_{ab}$ into  the lapse ($N$), shift ($N_\alpha$) and $3$-metric $h_{ab} = g_{ab} + u_au_b$. The comoving observers with velocity
$u^a$ will have the (in general, nonzero) acceleration  $a_i\equiv u^j\nabla_ju_i=h^j_i(\nabla_jN/N)$ which is purely spatial (i.e., $u^ia_i=0$) and has the magnitude $a\equiv\sqrt{a_ia^i}$.
The conditions $t(x)=$ constant, $N(x)=$ constant, taken together, define the 2-dimensional surface $\mathcal{S}$ (` equipotential surface') with the area element $\sqrt{\sigma}d^{D-2}x$ and the binormal $\epsilon_{ab}$ which we can take it to be $\epsilon_{ab}\equiv r_{[a}u_{b]}$  where  $r^\alpha=\pm(a^\alpha/a)$ is essentially the unit vector along the acceleration that points outwards.  
Comoving observers moving with velocity $u^a$ will have the acceleration $a$ which allows us to introduce the notion of a local Rindler frame at any event $\mathcal{Q}$ with this acceleration. A null surface passing though $\mathcal{Q}$ will now act as a patch of horizon to the local Rindler observers. They will attribute the (Tolman-corrected) Davies-Unruh \cite{du} temperature $T=Na/2\pi$ to the vacuum state of the freely falling observers. Further,  one quarter of the area element $dS=\sqrt{\sigma}d^2x/4 L_P^2$ can be thought of as the entropy associated with this patch of horizon in general relativity. (For more details of this construction and interpretation, see \cite{grtp, mpla}.)

While \eq{a3} associates a gravitational momentum $\gm q$ with \textit{any} vector field $q^a$, the momentum related to the  time evolution vector, $\xi^a \equiv N u^a$, is of special interest. This vector  measures the  proper-time lapse corresponding to the normal $u_a  = - N \nabla_a t$ to the  $t=$ constant surfaces. In static spacetimes,   $\xi^a$ can be chosen to be the timelike Killing vector. Since one motivation for the  Lie derivative $\pounds_q$ in \eq{a3}
is based on the idea of generalizing the time derivative occurring in $\dot p$, the vector $\xi^a$ can be a natural choice for the taking the Lie derivative \cite{grtp}. For this choice, $\gm \xi$ does have very interesting interpretation. 
We can now show that\cite{grtp,mpla} the on-shell \textit{total} energy in a region, bounded by $N=$ constant surface, is given by 
the thermal energy of the boundary, defined with Davies-Unruh temperature of local Rindler observers and the entropy density $s=(1/4)\sqrt{\sigma}$:
\begin{equation}
-\int_{\mathcal{V}} \sqrt{h}\, d^3 x \, u_a\, \mm \xi =  \int_{\partial \mathcal{V}} d^2 x \, Ts
\label{totE}
\end{equation}
(Note that the energy density is $-P^au_a$ with our signature convention.)
Further, the  the off-shell \textit{gravitational energy} in the same region is given (see e.g., Eq.(84) of \cite{grtp}) by:
\begin{equation}
-\int_\mathcal{R} d^3x \, \sqrt{h}\, u_a \gm \xi = \int_\mathcal{R} d^3x \, \sqrt{h}\,\mathcal{H}_{\rm adm} + \int_{\partial \mathcal{R}} d^2 x \, Ts
\end{equation} 
The first term on the right is the integral of  
$\mathcal{H}_{\rm adm} \equiv -N ( K^2 - K_{ab} K^{ab} + {}^3R)=-2NG_{ab}u^au^b $  
while the second term is the thermal energy of the boundary. This is an off-shell result.\footnote{It does not seem to have been widely appreciated that Noether  current $J^a[q]$ associated with an arbitrary vector field $q^a$ is, in general, \textit{non-zero in the flat spacetime} --- which, incidentally,  is yet another reason not to link it to diffeomorphism invariance of the Hilbert action. So the $\gm q=\mm q$ attributed to a \textit{flat spacetime} by, say, accelerated observers  can be nonzero. This is a feature and not a bug; and \eq{totE} relates it to the thermal effects seen even in the flat spacetime.}

When the spacetime  is static $\xi^a$ can be chosen to be the natural timelike Killing vector. In this case, we know that the vector $T^a_b \xi^b$ is indeed conserved. But the matter momentum defined as $-T^a_b u^b$ where  $u^a\equiv \xi^a/(-\xi^b\xi_b)^{1/2}$ is the four-velocity of an observer moving along the orbits of the Killing vector is \textit{not}, in general, conserved. (This is another motivation for using $\xi^a=Nu^a$ in general.)  It follows that, in any static geometry, the gravitational momentum $\gm{\xi^a}$ is also \textit{separately} conserved, which is understandable.

Finally, we mention that the gravitational momentum plays a crucial role in the thermodynamics of the null surface \cite{SKP,SP}. The projection of gravitational momentum along the normal to the null surface, and in the orthogonal directions, allows us to write the different components of Einstein's equations in an insightful manner. In particular, the flow of gravitational momentum along the null congruence defining the null surface leads to a thermodynamic identity of the form $TdS = dE + PdV$ where $P$ is the work function. Another projection orthogonal to the null surface allows two components of the field equations to be written in the form of the Navier-Stokes equation. These results show that the gravitational momentum is closely related to the thermal properties of the null surface which, in turn, form the corner stone of the emergent gravity paradigm.

If one is willing to accept the definition  of $\gm q$ in \eq{a3} as primary (or derive it from more fundamental considerations), then the dynamics has a nice description. One starts with an energy momentum tensor $T^a_b$ for matter and a gravitational momentum tensor $\gm q$ for gravity. Their conservation laws lead to equations of motion for matter and gravity!.

\section*{Acknowledgement}

I thank Sumanta Chakraborty, S. Date and D. Kothawala for comments on an earlier draft. My work is partially supported by the J.C.Bose research grant of the Department of Science and Technology, Government of India.


\begin{thebibliography}{99}

\bibitem{mtw} C. W. Misner, K. S. Thorne, and J. A. Wheeler, (1973),  \textit{Gravitation},  (W. H Freeman, USA).

\bibitem{KBP} Krishnamohan Parattu, Bibhas Ranjan Majhi, T. Padmanabhan,    \textit{Phys.Rev.,} \textbf{D 87},  124011 (2013) [arXiv:1303.1535].

\bibitem{grtp}  
T. Padmanabhan,  Gen.Rel.Grav, \textbf{46}, 1673 (2014) [arXiv:1312.3253].

\bibitem{mpla} T. Padmanabhan, \textit{ Mod.Phys.Letts} \textbf{A, 30}, 1540007 (2015) [arXiv:1410.6285]. 

\bibitem{ayan} A. Mukhopadhyay, T. Padmanabhan,  \textit{Phys.Rev.,} \textbf{D 74,} 124023 (2006) [hep-th/0608120] 

\bibitem{du} 
P. C. W. Davies,  \textit{J. Phys.}, \textbf{A 8},  609 (1975); 
W. G. Unruh,   \textit{Phys. Rev.} \textbf{D 14}, 870 (1976).

\bibitem{SKP} Sumanta Chakraborty, Krishnamohan Parattu, T. Padmanabhan, (2015) 
\textit{Gravitational Field equations near an Arbitrary Null Surface expressed as a Thermodynamic Identity},
[arXiv:1505.05297]

\bibitem{SP} Sumanta Chakraborty, T. Padmanabhan, (2015) \textit{in preparation}.


\end{thebibliography}
\end{document}